\documentclass[%
 reprint, 
superscriptaddress,
 amsmath,amssymb,
 aps,
pra, 
longbibliography, floatfix ]{revtex4-1}

\usepackage{natbib} 
\usepackage{amsmath} 
\usepackage{graphicx} 
\usepackage{longtable}
\usepackage{multirow}
\usepackage{amsmath}
\usepackage{epstopdf}
\usepackage{placeins}
\usepackage{MnSymbol}

\usepackage{xcolor}  

\usepackage[english]{babel}

\makeatletter
\def\bbl@set@language#1{%
  \edef\languagename{%
    \ifnum\escapechar=\expandafter`\string#1\@empty
    \else\string#1\@empty\fi}%
  \@ifundefined{babel@language@alias@\languagename}{}{%
    \edef\languagename{\@nameuse{babel@language@alias@\languagename}}%
  }%
  \select@language{\languagename}%
  \expandafter\ifx\csname date\languagename\endcsname\relax\else
    \if@filesw
      \protected@write\@auxout{}{\string\select@language{\languagename}}%
      \bbl@for\bbl@tempa\BabelContentsFiles{%
        \addtocontents{\bbl@tempa}{\xstring\select@language{\languagename}}}%
      \bbl@usehooks{write}{}%
    \fi
  \fi}
\newcommand{\DeclareLanguageAlias}[2]{%
  \global\@namedef{babel@language@alias@#1}{#2}%
} \makeatother

\DeclareLanguageAlias{en}{english}

\begin{document}

\title{Extending Modified Module Analysis to Include Correct Responses: An Analysis of the Force Concept Inventory}
\author{Jie Yang}
\affiliation{%
    West Virginia University, Department of Physics and Astronomy,
    Morgantown WV, 26506
}%

\author{James Wells}
\affiliation{%
    College of the Sequoias, Science Division, Visalia CA, 93277
}%
\author{Rachel Henderson}
\affiliation{%
    Michigan State University, Department of Physics and Astronomy,
    East Lansing MI, 48824
}%
\author{Elaine Christman}
\affiliation{%
    West Virginia University, Department of Physics and Astronomy,
    Morgantown WV, 26506
}%
\author{Gay Stewart}
\affiliation{%
    West Virginia University, Department of Physics and Astronomy,
    Morgantown WV, 26506
}%
\author{John Stewart}
\email{jcstewart1@mail.wvu.edu}
\affiliation{%
    West Virginia University, Department of Physics and Astronomy,
    Morgantown WV, 26506
}%

\date{\today}

\begin{abstract}

Brewe, Bruun, and Bearden first applied network analysis to
understand patterns of incorrect conceptual physics reasoning in
multiple-choice instruments introducing the Module Analysis for
Multiple-Choice Responses (MAMCR) algorithm. Wells {\it et al.}
proposed an extension to the algorithm which allowed the analysis
of large datasets called Modified Module Analysis (MMA). This
method analyzed the network structure of the correlation matrix of
the responses to a multiple-choice instrument. Both MAMCR and MMA
could only be applied to networks of incorrect responses. In this
study, an extension of MMA is explored which allows the analysis
of networks involving both correct and incorrect responses. The
extension analyzes the network structure of the partial
correlation matrix instead of the correlation matrix. The new
algorithm, called MMA-P, was applied to the FCI and recovered much
of the structure identified by MMA. The algorithm also identified
sets of correct answers requiring similar physical reasoning
reported in previous studies. Beyond groups of all correct and all
incorrect responses, some groups of responses which mixed correct
and incorrect responses were also identified. Some of these mixed
response groups were produced when a correct response was selected
for incorrect reasons; some of the groups helped to explain the
gender unfairness previously reported for the FCI.

\end{abstract}

\maketitle

\section{Introduction}

The structure of students' conceptual understanding of physics and
the evolution of that understanding has been an important research
strand within Physics Education Research (PER) since its
inception. Research into student understanding has often relied on
multiple-choice conceptual instruments such as the Force Concept
Inventory (FCI) \cite{hestenes1992} and the Force and Motion
Conceptual Evaluation \cite{thornton1998}. Quantitative research
examining these instruments began shortly after their introduction
\cite{huffman_what_1995}. Recently, a new class of quantitative
methods using network analysis has been applied to further
understand common student difficulties with mechanics
\cite{brewe2016,wells2019}.

Network analysis is a broad and powerful set of analytic
techniques applicable to situations as diverse as identifying
drought resistant genes in plants \cite{sircar_functional_2015}
and exploring the status hierarchies of teenagers
\cite{smith_movement_2015}. A network is formed of nodes and edges
where edges connect pairs of nodes. Edges can have numerical
weights that represent properties which influence the strength of
the relationship. Network analysis seeks to characterize the
structure of the network. One important type of structure is the
identification of groups of nodes that are more connected to each
other than they are to other nodes in the network. These subgroups
are called ``communities'' or ``modules'' interchangeably; we will
call them communities to conform with the naming conventions of
the ``igraph'' package \cite{igraph} in the ``R'' software system
\cite{R-software} which was used for this analysis.

A substantial body of research has investigated differences in the
conceptual performance of groups underrepresented in physics
\cite{wieman2019,hendersonperc2017}; the majority of this work has
examined differences between men and women \cite{madsen2013}. The
new network analytic techniques have been used to explore
differences in the structure of incorrect reasoning of men and
women \cite{wells2019}. We continue this practice and report
results disaggregating men and women.

\subsection{Research Questions}

The network analytic method used in this work extends Modified
Module Analysis (MMA) introduced by Wells {\it et al.}
\cite{wells2019} as an extension to the original algorithm
pioneered by Brewe, Bruun, and Bearden \cite{brewe2016}. Both
methods  were only productive when applied to the incorrect
answers in the FCI; the correct answers formed a single tightly
connected community that prevented identification of additional
structure. Modified Module Analysis is described in detail in Sec.
\ref{sec:mma}. The purpose of this study is to explore an
extension of MMA to include correct responses; this extension is
called MMA-P. This work was performed using the same dataset as
the previous MMA study; this dataset was also used by Traxler {\it
et al.} to demonstrate that a subset of the items in the FCI are
substantially unfair to women \citep{traxler2018}. This study will
use the results of the MMA-P algorithm to seek an explanation for
this unfairness.

This study explored the following research questions:
\begin{description}
\item[RQ1] How can Modified Module Analysis be extended to include
correct responses? Are the communities detected by the extended
algorithm productive in furthering the understanding of the FCI?
How are the communities identified by the extended algorithm
related to structures identified in previous studies? \item[RQ2]
How do the incorrect communities change between the pretest and
the post-test? \item[RQ3] How do the communities detected differ
between men and women?
\end{description}

\subsection{Prior Studies}
This work draws heavily on two prior studies of the FCI performed
on the same dataset as used in this study. These studies will be
referenced as Study 1 and Study 2 in this work.

\subsection{Study 1 : Modified Module Analysis}
\label{sec:mma} Brewe, Bruun, and Bearden \cite{brewe2016} first
introduced network analysis to PER as Module Analysis for
Multiple-Choice Responses (MAMCR). Their initial work was applied
to a sample containing only 143 FCI student responses. Wells {\it
et al.} \cite{wells2019} sought to replicate this study with a
large sample of FCI pretest and post-test responses
($N_{pre}=4509$ and $N_{post}=4716$) and found the original MAMCR
method did not scale to large datasets. To extend network analysis
to large datasets, they introduced a modification of the original
MAMCR algorithm called Modified Module Analysis (MMA). The Wells
{\it et al.} \cite{wells2019} study will be referenced as Study 1
in this work.

In MMA, the network is formed of nodes representing responses to
the FCI. For example, the network contains a node representing the
selection of response D to FCI item 2, node 2D. The edges in the
network connect nodes that are correlated above some threshold; in
Study 1, $r>0.2$ was used as the threshold where $r$ is the
correlation coefficient. The correlation coefficient is used as
the weight of the edge. For example, if responses 2D and 5E have a
correlation coefficient of $r=0.25$, then the network will contain
an edge between nodes 2D and 5E with weight $0.25$. For productive
analysis using MMA, nodes representing correct responses must be
removed. If these nodes were included, they formed a densely
connected community which obscured other structures.

Study 1 identified 9 groups of incorrect answers on the pretest
and 11 groups on the post-test. In most cases, these groups could
be explained as common misconceptions using the taxonomy of
misconceptions introduced with the publication of the FCI
\cite{hestenes1992} and refined by Hestenes and Jackson
\cite{fcitable}. A subset of the incorrect communities was the
result of the use of item blocks in the FCI. An item block is a
group of items that refers to the same physical system or are
written with a common stem. Most of the pretest and post-test
groups identified were the same for men and women. The groups
identified also had little relation to the gender unfair items
identified by Traxler {\it et al.} \cite{traxler2018}. Study 1
concluded that the differences in the structure of incorrect
physical reasoning between men and women did not explain the
gender differences reported for the FCI or the unfairness of the
items in the instrument.

Study 1 was recently replicated by Wells {\it et al.} for the FMCE
\cite{wells2020}. Incorrect responses representing the same
misconception were identified in the same community demonstrating
the algorithm is productive for examining instruments beyond the
FCI. The incorrect communities identified, however, were often
different for men and women both pre- and post-instruction as was
the change in the communities from the pretest to the post-test.

\subsection{Study 2: Multidimensional Item Response Theory and FCI}

In Study 2, Stewart  {\it et al.} \cite{stewart2018} used
constrained Multidimensional Item Response Theory (MIRT) to
develop a detailed model of the physical reasoning needed to
correctly solve the FCI. Study 2 identified a number of subgroups
of items in the FCI requiring similar physical reasoning for their
solution. These subgroups were used extensively in Study 1 because
the incorrect communities found by MMA were often well aligned
with the groups of items requiring similar correct reasoning
identified in Study 2. These subgroups were \{4, 15, 16, 28\},
\{5, 18\}, \{6, 7\}, and \{17, 25\}. Study 2 showed that the
practice of blocking items was generating correlations between
items not related to the physical principles tested and, as such,
retained only the first item in an item block. The item blocks in
the FCI are \{5, 6\}, \{8, 9, 10, 11\}, \{15, 16\}, and \{21, 22,
23, 24\}. While items \{25, 26, 27\} are not explicitly blocked,
both items 26 and 27 refer to item 25 and, therefore, these three
items should be treated as an item block.

Wells {\it et al.} application of MMA to the  FMCE
\cite{wells2020} also relied on a constrained MIRT analysis of the
instrument to identify items requiring similar physical reasoning
\cite{yang2019}. As in Study 1, incorrect answer communities were
often formed of items requiring the same correct reasoning.

\subsection{Results of Prior Research}

Prior research into the structure of the FCI was thoroughly
summarized in Study 2, research into the misconceptions measured
by the FCI in Study 1, and issues of the gender fairness and how
they relate to the FCI in Traxler {\it et al.} \cite{traxler2018}.
These fairly extensive research strands are summarized below;
readers interested in more detail are directed to these previous
works.

\subsubsection{Exploratory Analyses of the FCI}

Hestenes, Wells, and Swackhammer decomposed the concept of force
into six conceptual dimensions and provided descriptions of the
concepts each FCI item was intended to measure
\cite{hestenes1992}. However, attempts to extract this factor
structure using exploratory factor analysis (EFA), which uses
correlations between all items to select groups that measure the
same idea, proved unsuccessful. Huffman and Heller identified only
two of these six factors, Newton's 3rd law and ``Kinds of
Forces,'' in their principal component analysis of a sample
consisting of 145 high school students. The only factor identified
in a sample of 750 university students was kinds of forces
 \cite{huffman_what_1995}.

Scott, Schumayer, and Gray found that while only one factor
explained much of the variance in an EFA study of the FCI
post-tests of 2150 college students, 5 factors were needed for an
optimal model \cite{scott2012exploratory}. Scott and Schumayer
identified a similar, but not identical, 5-factor model when using
MIRT on a related dataset, confirming their model and suggesting
that traditional EFA and MIRT are complementary techniques
\cite{scott2015}. Semak {\it et al.} applied factor analysis to
the FCI pretest and post-test responses of 427 college students
and found that an optimal model included five factors for the
pretest and six for the post-test \cite{semak2017}.  Study 2
conducted an exploratory MIRT analysis of college post-test data
and reported that a 9-factor model was optimal \cite{stewart2018}.

\subsubsection{Gender and the FCI} Differences between the performance
of men and women on the FCI and Force and Motion Conceptual
Evaluation (FMCE) have been broadly reported with men
outperforming women by $13\%$ on pretests and $12\%$ on
post-tests. Many explanations have been advanced to explain the
these differences including differences in high school physics
class election \cite{edu2009,nces2015,ets2016}, cognitive
differences
\cite{maeda2013,halpern,hyde1988gender,hyde1990gender}, and
psychocultural factors including mathematics anxiety
\cite{else2010cross,ma1999meta}, science anxiety
\cite{mallow1982,udo2004,mallow2010}, and stereotype threat
\cite{shapiro2012}.

All the above explanations locate the source of the performance
differences with the students, a separate body of research has
shown multiple items in the FCI are unfair to women, and some to
men \cite{dietz_gender_2012,popp2011,mccullough_differences_2001}.
An item is unfair if men and women with the same general ability
with the material score differently on the item. Gender
differences in some FCI items have been reported for many years.
Recently Traxler {\it et al.} \cite{traxler2018} provided
convincing evidence that 5 FCI items, items 14, 21, 22, 23, and
27, were substantially unfair to women using samples from three
institutions. In their largest sample, Differential Item Function
(DIF) theory \cite{clauser1998} identified a total of eight items
with large DIF; six were unfair to women (12, 14, 21, 22, 23, 27)
and two to men (items 9 and 15). The problematic items had
generally, but not consistently, been identified in earlier work
\cite{dietz_gender_2012,popp2011,mccullough_differences_2001}.

\subsubsection{The Structure of Knowledge}

MAMCR, MMA, and MMA-P all attempt to identify structure within
student responses to a multiple-choice instrument. MAMCR and MMA
only investigate incorrect responses and, thus, attempt to find
coherent patterns of incorrect reasoning. Understanding common
difficulties shared by many students in learning Newtonian
mechanics has been an important area of research in PER since its
inception
\cite{viennot1979,trowbridge1981,caramazza1981,peters1982,mccloskey1983,gunstone1987,camp1994}.
This research led to a systematic exploration of students'
understanding of Newton's laws and its epistemological development
\cite{mcdermott1997,thornton1998,rosenblatt2011,erceg2014,waldrip2014}.

From these investigations and research outside of PER, theoretical
frameworks of the structure of students' conceptual knowledge of
physics emerged. Within PER, patterns of incorrect answering have
often been characterized as ``misconceptions,'' consistently
applied incorrect reasoning specific to the physical concept
tested. Other theories have been advanced to explain incorrect
reasoning; two of the most prominent are the knowledge-in-pieces
framework \cite{disessa1993,disessa1998} and the ontological
categories framework \cite{chi1993,chi1994,slotta1995}.

The knowledge-in-pieces framework models student knowledge as
composed of small pieces of reasoning that are applied either
independently or collectively to solve a problem. Models
involveing small fragments of reasoning have been advanced by many
authors and have been called phenomenological primitives (p-prims)
\cite{disessa1993,disessa1998}, resources
\cite{hammer1996misconceptions,hammer_more_1996,hammer2000student},
and facets of knowledge \cite{minstrell1992}. In the
knowledge-in-pieces framework, what PER has called misconceptions
are the result of problems activating the same collections of
inappropriately applied knowledge pieces. In this framework,
misconceptions are not ``resolved,'' rather students' frameworks
must be refined to apply to the situation in the problems. The
ontological categories model suggests that incorrect student
answers are a result of misclassification of a physical concept.
For example, force may be misclassified as a substance.

Scherr \cite{scherr2007} provides a concise definition of both the
misconception view and the knowledge-in-pieces view which we will
adopt for this work. The misconception view is ``a model of
student thinking in which student ideas are imagined to be
determinant, coherent, context-independent, stable, and rigid''
\cite{scherr2007}. The knowledge-in-pieces framework models
student conceptual ideas ``as being at least potentially
truth-indeterminate, independent of one another,
context-dependent, fluctuating, and pliable'' \cite{scherr2007}.

The research in the current work most closely aligns with the
misconception and knowledge-in-pieces views. Study 1 and Wells
{\it et al.} \cite{wells2020} both observed that communities of
incorrect answers are generally formed of items requiring the same
physical reasoning for their solution and shared a common
misconception from the taxonomy of Hestenes and Jackson
\cite{fcitable}. Because the incorrect reasoning is applied across
multiple similar problems, the misconception view may be more
appropriate for the incorrect answer communities identified by MMA
or MMA-P. The FCI was developed partially to probe common
misconceptions \cite{hestenes1992}, so the finding that this
framework is more appropriate is not a surprise. Network analysis
could potentially identify p-prims as consistently applied
reasoning divorced from the physical context, but thus far this
type of community has not been observed.

\section{Methods}

\subsection{Instrument}

The FCI is a 30-item instrument designed to assess a student's
conceptual understanding of topics in Newtonian mechanics
including one- and two-dimensional kinematics and Newton's laws.
The FCI's coverage of Newtonian mechanics is not exhaustive; the
instrument contains no questions on common topics from
introductory mechanics courses such as conservation of momentum
and energy \cite{hestenes1992}. Each item includes four incorrect
responses with distractors intended to elicit common student
misconceptions as well as one correct response.  This study uses
the revised FCI released in 1995 \cite{fci-revised}, which is
available at PhysPort \cite{physport}.

\subsection{Sample}

\label{sec:samples} The data for this study were collected at a
large southern land-grant university serving approximately 25,000
students. Overall university undergraduate demographics were 79\%
white, 5\% African American, 6\% Hispanic, with other groups each
3\% or less for the period studied \cite{usnews}.

The sample consists of 4509 complete pretest records (3482 men and
1027 women) and 4716 complete post-test records (3628 men and 1088
women) from students enrolled in an introductory calculus-based
mechanics course. The class served primarily engineering and
physical sciences majors and was presented by the same lead
instructor using a consistent pedagogy throughout the study. This
instructor implemented interactive engagement activities in the
lecture and multiple research-based instructional methods were
used in the required laboratory sections.

\subsection{Partial Correlation}
To allow the investigation of correct and incorrect answers
together in the same network, the correlation matrix used by MMA
was replaced by the partial correlation matrix. Partial
correlation measures the association between two variables while
eliminating the effect of one or more other variables. The
correlation, $r_{XY}$, between variable $X$ and variable $Y$ is
defined in Eqn. \ref{eq:corr}
\begin{equation}
\label{eq:corr} r_{XY}=\frac{E[(X-\mu_X)(Y-\mu_Y)]}{\sigma_X
    \sigma_Y}
\end{equation}
where $\mu_i$ is the mean of variable $i$, $\sigma_i$ is the
standard deviation, and $E[X]$ is the expectation value.

Two variables may be correlated because they are both related to a
third variable; the partial correlation controls for the relation
with a third variable. The partial correlation of variable $X$ and
variable $Y$ controlling for variable $Z$, $r_{XY|Z}$, is defined
in Eqn. \ref{eq:pcor}.
\begin{equation}
\label{eq:pcor} r_{XY|Z}=\frac{r_{XY}-r_{XZ}r_{YZ}}{\sqrt{1-r_{XZ}^2}\sqrt{1-r_{YZ}^2}}
\end{equation}

Conceptually, partial correlation can best be understood in terms
of linear regression. If $X$, $Y$, and $Z$ are continuous random
variables and $\epsilon_{XZ}$ are the residuals when $Z$ is
regressed on $X$ (controlling $X$ for $Z$) and $\epsilon_{YZ}$ are
the residuals when $Z$ is regressed on $Y$, then the partial
correlation $r_{XY|Z}$ is the correlation of the residuals,
corr($\epsilon_{XZ}$, $\epsilon_{YZ}$). The residuals are the part
of the variance of $X$ or $Y$ not explained by the variation of
$Z$.

The package ``ppcor'' \cite{kim2015ppcor}, part of the R software
system \cite{R-software}, was used to perform the partial
correlation analysis.

\subsection{Extending Modified Module Analysis}

Modified Module Analysis uses community detection algorithms on
the network defined by the correlation matrix of all incorrect
responses with a threshold applied; Study 1 used $r>0.2$, as will
this study. Correlation matrix entries below this threshold are
set to zero. The non-zero entries represent edges in the network;
the edge weight is the strength of the correlation. Study 1
applied a number of additional filters to eliminate insignificant
edges and rarely selected nodes; these were repeated. Nodes
selected by fewer than 30 students were removed. Edges connecting
different responses to the same item (and thus strongly negatively
correlated) were removed; the $r>0.2$ threshold naturally removed
these edges. Study 1 also removed edges that were not
statistically significant after a Bonferroni correction was
applied. At the sample size of this study and Study 1, this
removed no additional edges because all edges with $r>0.2$ were
significant.

To allow MMA to productively identify communities when the correct
response nodes are included in the network, the correlation matrix
was replaced with the partial correlation matrix controlling for
total FCI score. This modified algorithm is called Modified Module
Analysis - Partial (MMA-P). Study 2 presented both the correlation
matrix and the partial correlation matrix of the correct answers
to FCI for this sample. The two matrices were very different with
the correlation matrix forming a sparsely connected network and
the partial correlation matrix separating into small sub-networks.

The partial correlation matrix has some theoretical advantages
over the correlation matrix if the goal is to identify
consistently applied reasoning. Two items in an instrument can be
correlated because either only the highest performing students get
them correct or because only the lowest performing students get
them wrong; they are correlated due to overall test score. To
remove these correlations, the partial correlation matrix controls
for overall test score. This explains why in Study 2 the
correlation matrix was generally connected with positive
correlations, but the partial correlations contained isolated
communities; the correlations between items in the FCI are
generally related to the total FCI score.

The MMA algorithm was then applied to the partial correlation
matrix. The fast-greedy community detection algorithm (CDA) was
applied to the network defined by the partial correlation matrix;
this identified subsets of the network which were more connected
to each other than they were to nodes outside the community. The
community structure is sensitive both to random noise in the data
and to the stochastic nature of CDAs. To determine the structure
not related to these factors, the dataset was bootstrapped with
1000 replications; the CDA was run on each replication. The R
package ``boot''  was used for bootstrapping \cite{boot1}. As in
Study 1, the dataset was downsampled to provide a sample balanced
between men and women. The number of times each pair of responses
was found in the same community was recorded for each of the 1000
samples forming a ``community matrix.'' The community fraction,
$C$, was defined as the fraction of the bootstrap replications in
which node i and j were in the same community. Only nodes found in
$C>80\%$ of the communities were analyzed.

\section{Results}

\def\xbox{$\times$}
\def\mrg{$\odot$}
\def\msn{$\bigcircle$}
\def\bth{$\otimes$}

\begin{table*}[!htb]
    \caption{Communities identified in the pretest and post-test at $r>0.2$ and
    $C>80\%$. The number in parenthesis is the
        intra-community density, $\gamma$,
        for communities where the intra-community density is not one. Communities labeled \bth\ were identified in both Study 1 and the current study. Communities labeled \msn\
        were identified in Study 1 but not in the current study.
        Communities labeled \xbox\  were identified in the current study, but not in Study 1. Communities labeled \mrg\ were identified in Study 1 as completely
        incorrect communities, but combined with correct responses to form mixed correct-incorrect communities in the current study.\label{tab:commat} }
    \centering
    \begin{tabular}{|l|cc|cc|l|}
        \hline
        \multirow{2}{*}{Community}&\multicolumn{2}{|c|}{Pretest}&\multicolumn{2}{|c|}{Post-test}&\multirow{2}{*}{Misconception/Principle/Explanation}\\
        &Men&Women&Men&Women&\\\hline
        \multicolumn{6}{|c|}{Completely Incorrect Communities}\\\hline
        1A, 2C&\mrg&&           &&Heavier objects fall faster.\\\hline
        1D, 2D&\bth&&           &&Lighter objects fall faster.\\\hline
        4A, 15C, 28D&\bth&\bth& \mrg    &\mrg&Newton's 3rd law misconceptions.\\\hline
        5D, 18D&&&            \bth&\xbox&Motion implies active forces.\\\hline
        5D, 11C, 13C, 18D, 30E&&&&\msn&Motion implies active forces.\\\hline
        5E, 18E&\bth&\bth&            \bth&\msn&Motion implies active forces/Centrifugal force.\\\hline
        6A, 7A&\msn&\bth&          \bth&\bth&Circular impetus.\\\hline
        8A, 9B&\bth&\bth&          \bth&\bth&Blocked item.\\\hline
        11B, 29A&\bth&\msn&        &&Motion implies active forces.\\\hline
        11C, 13C&&&       \xbox&\xbox&Motion implies active forces.\\\hline
        11C, 13C, 30E&&&       \msn&&Motion implies active forces.\\\hline
        15D, 16D&\xbox&&       &&Newton's 3rd law misconceptions.\\\hline
        17A, 25D&&&           \bth&\msn&Largest force determines motion.\\\hline
        21B, 23C&&&           \bth&\bth&Blocked item.\\\hline
        21C, 22A&\bth&&  \mrg       &\mrg&Blocked item.\\\hline
        23D, 24C&\bth&\bth&         \bth&\bth&Impetus dissipation.\\\hline

        \multicolumn{6}{|c|}{Mixed Correct and Incorrect Communities}\\\hline

        1A, 2C, 17B*&\xbox&&           &&Unknown\\\hline
        4A, 14D*, 15C, 28D&&&     &\xbox(67\%)&Newton's 3rd law and 2D kinematics.\\\hline
        4A, 15C, 21E*, 22B*, 28D&&&     \xbox&&Newton's 3rd law and 2D kinematics.\\\hline
        8B*, 9E*, 15A*, 21C, 22A&&&         &\xbox&8B and 21C share a similar trajectory; 8B and 9E are blocked.\\\hline
        8B, 21C*&&\xbox&         &&8B and 21C share a similar trajectory.\\\hline
        8B*, 21C, 22A&&&         \xbox&&8B and 21C share a similar trajectory.\\\hline
        23A,  24A*&&\xbox&         &&24A is the correct answer if 23A were correct.\\\hline

        \multicolumn{6}{|c|}{Completely Correct Communities}\\\hline

        4E*,28E*&&&       &\xbox&Newton's 3rd law.\\\hline
        4E*, 15A*, 28E*&&\xbox(67\%)&\xbox       &&Newton's 3rd law.\\\hline
        5B*, 18B*&&&            \xbox&\xbox&Centripetal acceleration in a curved trajectory. \\\hline
        6B*, 7B*&&&          \xbox&\xbox&Instantaneous velocity is tangent to the trajectory.\\\hline
        11D*, 13D*&&&       &\xbox&Motion under gravity; a force in the direction of motion is not necessary. \\\hline
        15A*, 28E*&\xbox&&       && Newton's 3rd law.\\\hline
        17B*, 25C*&&\xbox&           &\xbox&Newton's 1st law; Addition of forces.\\\hline
        \multirow{2}{*}{17B*, 25C*, 26E*}&&&          \multirow{2}{*}{\xbox}&& Newton's 1st and 2nd law; Addition of forces;\\
        &&&          && (26E) 1D acceleration.\\\hline
       21E*, 22B*, 26E*&&&           &\xbox&Newton's 2nd law; 1D and 2D kinematics.\\\hline

    \end{tabular}
\end{table*}
\begin{figure*}[!htb]
    \centering
    \includegraphics[width=7in]{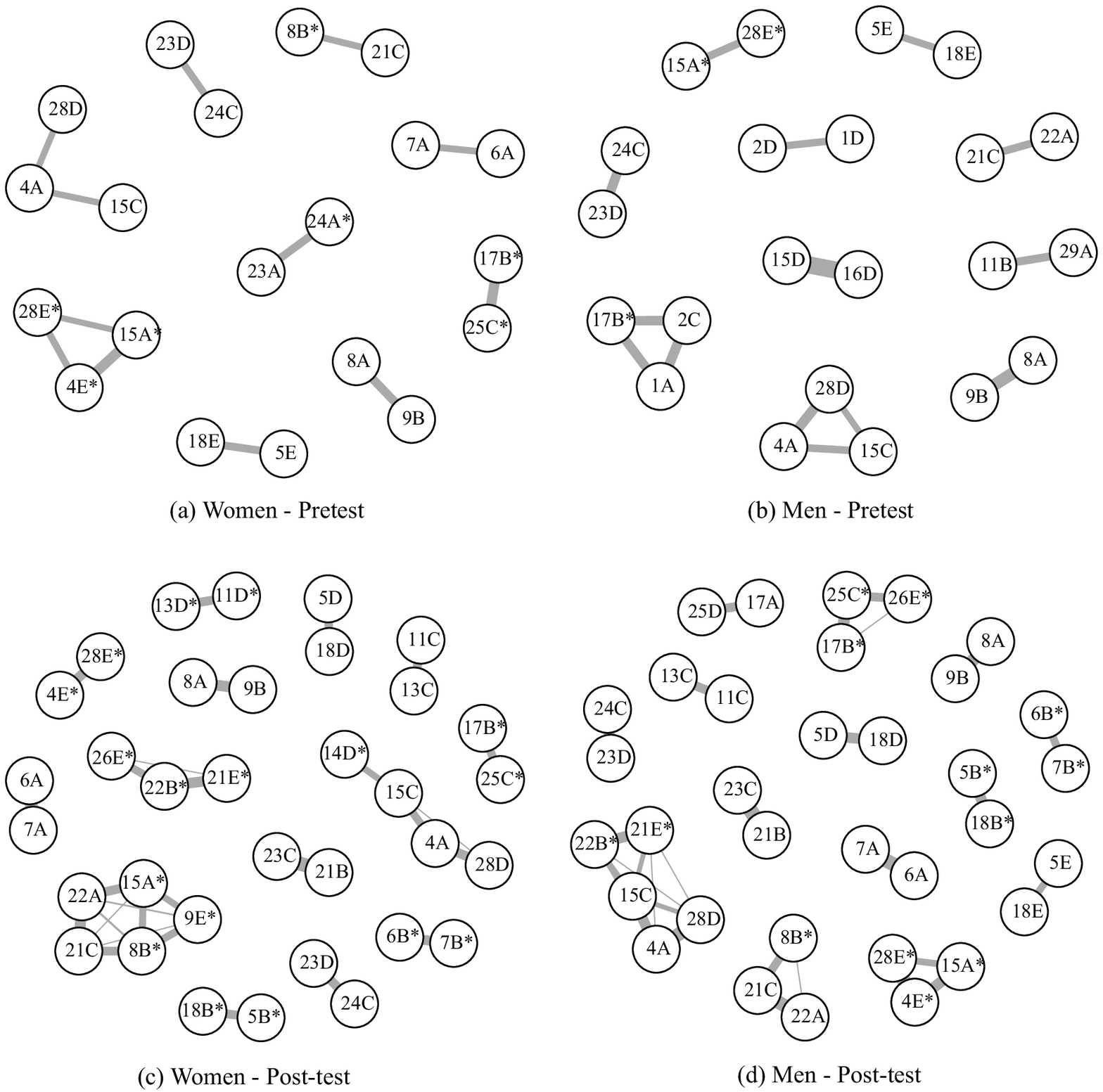}
    \caption{Communities detected in the FCI partial correlation matrix with
    $r>0.2$. The strength of the correlation is represented by the
    line thickness.
    \label{fig:cor-com}}
\end{figure*}
\subsection{Extended Modified Module Analysis}

Modified Module Analysis - Partial was applied to understand the
structure of the FCI. The communities detected with a partial
correlation threshold of $r>0.2$ identified in 80\% of the
bootstrap replications ($C>80\%$) are shown in Table
\ref{tab:commat}. The communities are represented graphically in
Fig. \ref{fig:cor-com}. Nodes representing correct responses are
identified with an asterisk (*). The communities naturally divide
into three classes as shown in Table \ref{tab:commat}: communities
formed of only correct answers, communities formed of only
incorrect answers, and communities that mix correct and incorrect
answers. Each type of community is examined below. Some
communities identified by MMA-P were also identified in Study 1
and are marked \bth, some were only identified in this study and
are marked \xbox, some were only identified in Study 1 and are
marked \msn; some communities identified in Study 1 merged with
correct answers to formed mixed correct-incorrect communities and
are marked \mrg.

Figure \ref{fig:cor-com} is a refinement of the network
visualization used in Study 1 and in Wells {\it et al.}
\cite{wells2020}. In these previous studies, a line was drawn
between the two nodes if the nodes were identified in the same
community 80\% of the time ($C>80\%$); however, two nodes can be
found in the same community because they are always connected to a
third node. For example, in Fig. \ref{fig:cor-com}(c) responses
\{21E*, 22B*, 26E*\} were detected in the same community in at
least 80\% of the replications; however, 21E* and 26E* are weakly
correlated. Both are strongly correlated with 22B*. To provide a
better indication of the degree of connection of two nodes in the
individual bootstrap replications of the CDA, the average
correlation of the two nodes across the 1000 replicates is used as
the edge weight in Fig. \ref{fig:cor-com}.

The degree of connection of a community is characterized by the
intra-community density, $\gamma$, the ratio of the number of
edges observed to the maximum possible number of edges
\cite{zweig}. For example, a community which contains four nodes
can be connected with a maximum of six edges. If only four of
those edges are observed, then $\gamma=4/6$. For communities that
are not fully connected ($\gamma<1$), $\gamma$ is presented in
parenthesis in Table \ref{tab:commat}.

\subsection{The Structure of the Response Communities}
\subsubsection{Completely Incorrect Communities}
The original MMA algorithm could only explore incorrect responses,
and therefore, all communities reported in Study 1 were composed
of incorrect responses. The incorrect communities identified by
MMA and MMA-P were very similar but not identical.

Study 1 showed that there were two classes of incorrect
communities. One was related to the practice of blocking items;
for other communities, the students seemed to be applying
consistent incorrect reasoning, a misconception.

Communities \{8A, 9B\}, \{21B, 23C\}, and \{21C, 22A\} are answers
within blocked problems where the second answer in the pair would
be correct if the first answer was correct. These were fairly
consistently detected by both MMA and MMA-P; however, the \{21C,
22A\} community merged with some correct responses to form a mixed
community post-instruction for both men and women in MMA-P. This
occurred because, post-instruction, the community joined with
response 8B* for men and responses 8B*, 9E*, and 15A for women.
These new communities are discussed with the mixed correct and
incorrect communities.

Three incorrect communities were identified by MMA-P, but not MMA.
In community \{15D, 16D\}, both items are incorrect responses to
the Newton's 3rd law group identified by Study 2; however, they
are also part of an item block. It is, therefore, impossible to
determine if the community results from a shared misconception or
from the effect of blocking. Item 16 is inconsistently identified
with other Newton's 3rd law items in exploratory factor analysis
studies
\cite{stewart2018,huffman_what_1995,hestenes1995interpreting,heller1995interpreting,
scott2012exploratory,semak2017,scott2015}. The two other incorrect
communities identified by MMA-P but not MMA resulted from
splitting the \{5D, 11C, 13C, 18D, 30E\} community into the
communities \{5D, 18D\} and \{11C, 13C\}.

\subsubsection{Completely Correct Communities}

Study 2 identified four groups of items requiring similar solution
structure: \{4, 15, 16, 28\}, \{5, 18\}, \{6, 7\}, and \{17, 25\}.
Three of the completely  correct communities combine items in the
first group,  \{4E*, 15A*, 28E*\}, \{4E*, 28E*\}, and \{15A*,
28E*\}; all of which require Newton's 3rd law for their solution.
All items in the three-item communities are strongly correlated in
Fig. \ref{fig:cor-com}.

Correct communities \{5B*, 18B*\}, \{6B*, 7B*\}, and \{17B*,
25C*\} were all identified as requiring similar reasoning for
their solution in Study 2. We note the prevalence of the B
response is a result of the extremely unbalanced use of
distractors in the FCI \cite{devore_examining_2016}. The correct
community \{17B*, 25C*, 26E*\}  extends the community \{17B*,
25C*\} by adding item 26. Both items 17 and 25 require the
addition of forces and Newton's 1st law for their solution. Item
26 requires the addition of forces, but the forces are unbalanced,
requiring Newton's 2nd law and one-dimensional kinematics (a net
force in the direction of motion causes an object to speed up).
Because the correct solution structure is substantially different,
it seems likely this community is detected because items 25 and 26
are in the same item block. This is supported by Fig.
\ref{fig:cor-com}(d) where there is a weak correlation between
responses 17B* and 26E* in the community \{17B*, 25C*, 26E*\}.

Both items in the correct community \{11D*, 13D*\} ask the
students to identify the forces on an object in motion; gravity
and a normal force for item 11 and gravity alone for item 13. Both
also explicitly test the motion implies active forces
misconception. Item 11 was eliminated from the constrained MIRT
analysis in Study 2 because of blocking and, therefore, no item
level measure of the discrimination of the item on the use of the
normal force is available. With the identification of the correct
community, \{11D*, 13D*\}, it seems likely that both items largely
test knowledge of the existence of a downward force of gravity
while in motion.

The correct community \{21E*, 22B*, 26E*\} contained blocked items
21 and 22 along with item 26. Items 21 and 22, while blocked, test
fairly different physical principles; however, items 22 and 26
both require the principle that if a force is applied in the
direction of motion, an object will speed up. There are strong
correlations between responses 21E* and 22B* and between responses
22B* and 26E* in Fig. \ref{fig:cor-com}(c), but a weak correlation
between 21E* and 26E*. It seems likely blocking caused a relation
of items 21 and 22 and the shared principle produced the relation
between items 22 and 26.

\subsubsection{Mixed Correct and Incorrect Communities}

In addition to examining the grouping of incorrect or correct
items, MMA-P can also detect communities combining correct and
incorrect items. These communities can provide further insight
into the functioning of the FCI and the complexities of student
thinking.

Many of the mixed communities seem to form because a pair of
questions mixes a correct and an incorrect community. For example,
responses \{8B*, 21C\} both represent the same trajectory (a
parabolic curve); this trajectory is the correct answer to item 8
and the incorrect answer to item 21. This similarity serves to
explain the mixed communities \{8B*, 21C\} and \{8B*, 21C, 22A\}.
Items 21 and 22 are blocked items where response 22A is the
correct response if response 21C had been the correct response.
Responses 8B* and 22A are weakly related in Fig.
\ref{fig:cor-com}(d) making it more likely that the shared
trajectory is the cause of the community. If a student is
selecting a specific trajectory using both the correct and
incorrect reasoning, it may indicate that response 8B* is being
selected correctly without a solid understanding.

The relation of 8B* and 21C partially explains the community
\{8B*, 9E*,15A*, 21C, 22A\}; items 8 and 9 are blocked and to
answer item 9 correctly requires a correct response to item 8. It
is unclear why Newton's 3rd law item 15 is associated with this
group. Some items in this group are weakly correlated as shown in
Fig. \ref{fig:cor-com}(c), but the strong relation of 15A* to 8B*,
9B*, and 22A does not have a grounding in either the correct
solution structure of the items or a shared misconception. It is
also difficult to identify a p-prim that might have been activated
in each case. Items 21 and 22 were identified as substantially
unfair to women in Traxler {\it et al.} \cite{traxler2018}. The
relation of the correct response 15A* to 22A and correct response
8B* to 21C correcting for overall test score provides evidence
that these items are being incorrectly answered for reasons not
completely related to Newtonian reasoning ability.

Responses 21E* and 22B* are part of an item block. The
identification of these items in a community of incorrect Newton's
3rd law responses, \{4A, 15C, 28D\} for men, may indicate that men
are selecting the items correctly for the wrong reasons which
would partially explain the unfairness of items 21 and 22 reported
by Traxler {\it et al.} \cite{traxler2018}.

It is possible a similar effect explains the community \{4A, 14D*,
15C, 28D\}. If students are selecting correct response 14D* for
incorrect reasons, it could be more correlated with the incorrect
Newton's 3rd law items than would be predicted by the overall FCI
score. This community was only identified for women
post-instruction. Item 14 was one of the items identified as
substantially unfair to women in Traxler {\it et al.}
\cite{traxler2018}; if it is also being answered correctly by
women for incorrect reasons, it may be more unfair than previously
reported.

Responses 23A and 24A* are also part of an item block. Item 23A
tests the misconception of impetus dissipation where after a force
is removed the object returns to its original trajectory before
the force was applied. Item 24A* (constant speed) is the correct
answer for the correct reasoning for item 23, but it is also the
correct answer if the misconception of impetus dissipation
actually applied. This suggests that some students are getting
item 24 correct for the wrong reasons, and that the item block may
need to be restructured.

The community \{1A, 2C, 17B*\} all involve items with an object
moving under the force of gravity. There is little additional
relation between the items. It is difficult to make a theoretical
case for this community either because of a shared misconception
or the correct answer structure. It may be that these items,
identified as a community only for men on the pretest, use less
consistent reasoning better modeled using the knowledge-in-pieces
framework as a p-prim. Response 1A might be answered by applying
the ``heavier is faster'' principle, response 2C the ``heavier
travels farther'' principle, and response 17B* the ``constant
implies equal'' principles; all examples of a fragment of
reasoning that ``like implies like.''

\section{Discussion}
\subsection{Research Questions}
This study sought to answer three research questions; they will be
addressed in the order proposed.

{\it RQ1: How can Modified Module Analysis be extended to include
correct responses? Are the communities detected by the extended
algorithm productive in furthering the understanding of the FCI?
How are the communities identified by the extended algorithm
related to structures identified in previous studies?}

Modified Module Analysis was extended to include correct answers
by replacing the correlation matrix with the partial correlation
matrix correcting for overall FCI test score. The modified
algorithm was called Modified Module Analysis - Partial (MMA-P).
This modification allowed the identification of a number of
relatively small communities of correct and incorrect answers
producing a network quantitatively similar to the MMA algorithm,
but including correct answers.

Study 1 could only report completely incorrect communities. Table
\ref{tab:commat} provides a comparison between the two studies. Of
the 38 communities identified either on the pretest or post-test
for men or women, 24 were identified in both studies, 11 were
identified only in Study 1, and 3 only in the present study. Four
of the communities identified in only Study 1 merged with correct
answers to form mixed communities in this study. As such, MMA and
MMA-P produce fairly similar incorrect answer communities. The
remaining incorrect communities identified in Study 1, but not in
the present study, may have resulted from correlations with
overall test score where only very weak or very strong students
selected the items.

Until this study, module analysis was not productive for finding
communities of correct answers in the FCI. Study 2 presented both
the FCI correlation matrix and the partial correlation matrix for
the correct answers to the FCI for the dataset used in this study.
The correlation matrix was sparsely and randomly interconnected
and showed little community structure. Clear community structure
was evident in the partial correlation matrix. As such, it was not
surprising that modifying MMA to use the partial correlation
matrix allowed the extraction of compact correct answer
communities.

Study 2 did not use network analysis to form communities of
correct answers, but it did present a taxonomy of isomorphic and
blocked problems that can be compared with the community
structure. Two items are isomorphic if they required the same
reasoning process for their solution. Many of the components of
the correct communities in Table \ref{tab:commat} were identified
in Study 2 as having similar solution structure \{4E*, 15A*,
28E*\}, \{5B*, 18B*\}, \{6B*, 7B*\}, and \{17B*, 25C*\}. The
students do not consistently integrate the Newton's 3rd law items,
\{4E*, 15A*, 28E*\}, with different combinations observed for men
and women either on the pretest or post-test. The items in
community \{11D*, 13D*\} differ only by the addition of the normal
force; both items specifically test the motion implies active
forces misconception by providing a distractor which indicates
there is a force in the direction of motion. The incorrect
community, \{11C, 13C\}, of students who apply this misconception
was also identified. The addition of the normal force does not
seem an important discriminating factor in whether the items are
answered correctly.

The attachment of 26E* to the \{17B*, 25C*\} community of
isomorphic items for men likely resulted from the blocking of
items 25 and 26; the solution structure of item 26 does not
suggest it should be directly associated with the other two items.
The correlations in Fig. \ref{fig:cor-com} support this
interpretation.  Item 26E* is also associated with the blocked
items {21B*, and 22B*} for women; however, the association is much
stronger with item 22B*; both item 22 and 26 require application
of the principle that forces in the direction of motion cause
objects to speed up.

As such, the correct community structure is largely what was
suggested by Study 2 with some additional sets of items that are
being answered consistently which were not modeled as having
similar structure in Study 2. Some of the connections identified
continue to support the negative impacts of blocking items on the
interpretability of the results of the instrument.

No prior studies have examined community structure combining
incorrect and correct responses in the same community. The mixed
communities identified in this study seem to originate largely
from combinations of items that may not be working as intended in
the instrument; items that are being answered correctly for the
wrong reason or items that are possibly being misinterpreted
contributing to instrumental unfairness.

{\it RQ2: How do the incorrect communities change between the
pretest and the post-test? RQ3: How do the communities detected
differ between men and women?} The differences from pretest to
post-test differ for men and women, and therefore, these two
questions will be taken together. Table \ref{tab:commat} shows the
communities identified by MMA-P. The completely correct,
completely incorrect, and mixed correct-incorrect communities are
discussed independently.

For completely incorrect communities, incorrect communities
identified in Study 1 which merged with correct answers to form
mixed communities in this study (labeled \mrg\ in Table
\ref{tab:commat}) will be included in the incorrect communities;
the incorrect items in the community were still detected in the
same community. For men, 9 communities were detected on the
pretest and 10 on the post-test; only 5 were consistent from
pretest to post-test. For women, 5 communities were identified on
the pretest and 8 on the post-test; 4 were consistent between the
pretest and post-test. As such, women have more consistent
incorrect reasoning patterns between pretest to post-test, but men
have more communities of incorrect reasoning both preinstruction
and post-instruction.

Comparing men and women, only 4 of the 10 pretest communities were
consistent for men and women, while 8 of the 10 post-test
communities were consistent between men and women. As such, there
is little consistency in incorrect reasoning between men and women
on the pretest, but substantial consistency on the post-test. Men
and women also had a similar number of items transition from
completely incorrect in Study 1 to mixed correct-incorrect in this
study.

Six incorrect communities identified by MMA in Study 1 were not
identified by MMA-P in this study. These missing communities were
often related to new communities identified by MMA-P but not MMA.
The community \{11C, 13C, 30E\} identified by MMA became the
community \{11C, 13C\} using MMA-P for men on the post-test.
Response 11C and 13C test the force in the direction of motion
misconception while response 30E asks about the ``force of a
hit;'' as such, the dissolution of the community is
understandable. The community \{5D, 11C, 13C, 18D, 30E\}
identified by MMA became the communities \{5D, 18D\} and \{11C,
13C\} in MMA-P; both can be theoretically justified using the
framework of Study 2 and both test similar misconceptions in the
framework of Hestenes and Jackson \cite{fcitable}. It may be the
the identification of the \{5D, 11C, 13C, 18D, 30E\} was a result
of correlations generated by overall test score.

Four of the communities identified by MMA but not MMA-P were more
difficult to explain. Communities \{6A, 7A\}, \{5E, 18E\}, and
\{17A, 25D\} contain incorrect responses to items identified in
Study 2 as having the same correct solution structure. It is
unclear why the incorrect reasoning for these items is not applied
consistently. The correct answer communities for these items were
consistently identified post-instruction. It may be that students
are transitioning to correct reasoning and that the misconception
is no longer consistently applied. Pretest community \{11B, 29A\}
also was no longer identified by women; students giving responses
in this community fail to account for the normal force. It is
unclear why women do not do so consistently. Item 29 was
identified by Traxler {\it et al.} \cite{traxler2018} as having
poor psychometric properties, so it may be that the item is simply
not functioning well.

Few mixed communities were detected either preinstruction or
post-instruction for men or women; there were no consistent mixed
communities either from pretest to post-test or for men and women.

Only three correct communities were detected on the pretest for
either men or women; only one was also identified
post-instruction. Most correct communities were detected
post-instruction; four for men and six for women; only two of
these communities were consistent between men and women. One of
the inconsistent correct communities \{17B*, 25C*, 26E*\} likely
resulted from blocking items 25 and 26; if response 26E* were
removed the community \{17B*, 25C*\} would also be consistent
between men and women. Three of the other inconsistent communities
result from the inconsistent application of Newton's 3rd law.
Responses 4E*, 15A*, and 28E* were identified in the same
community and were highly correlated (Fig. \ref{fig:cor-com}(d))
for men on the post-test. Only responses 4E* and 28E* were in the
same community for women; response 15A* was identified as part of
a mixed community for women. If the minor differences in the
\{4E*, 15A*, 28E*\} and \{17B*, 25C*\} communities are ignored,
then all correct communities identified for men post-instruction
were also identified for women; two additional correct communities
were identified for women. The items in one of the additional
communities, \{11D*, 13D*\}, differ by the inclusion of a normal
force; this difference affects men but not women. The last
inconsistent community \{21E*, 22B*, and 26E*\} involves the
blocked items 21 and 22, but also item 26 which requires somewhat
different physical reasoning. It is unclear why these items are
identified in the same community; however, items 21 and 22 were
two items identified as highly unfair to women in Traxler {\it et
al.} \cite{traxler2018}.

\subsection{Other Observations}

The MMA-P algorithm identified much of the incorrect structure
identified by MMA while also productively identifying correct
structure. The structures identified were consistent with the
theoretical framework provided by Study 2 and with the taxonomy of
misconceptions of Hestenes and Jackson \cite{fcitable}. The mixed
correct-incorrect communities allowed the identification of
combinations of responses where symmetric correct and incorrect
reasoning was applied suggesting the correct response was being
selected for the incorrect reason. These communities also provide
some hint as to items within the FCI which were not functioning
correctly where correct answers were related to a community of
consistently applied misconceptions. In general, the MMA-P
extension of MMA provided a richer picture of the instrument.

\section{Future} The MMA-P algorithm will be applied to other
popular conceptual instruments such as the the FMCE, the
Conceptual Survey of Electricity and Magnetism \cite{maloney2001},
and the Brief Electricity and Magnetism Assessment \cite{ding2006}
to further understand their structure. The algorithm will also be
applied to multiple samples taken from students at different
institutions to understand how the incorrect reasoning structures
identified change with institutional setting.

\section{Conclusion}

This work extended the Modified Module Analysis algorithm of Wells
{\it et al.} \cite{wells2019} to allow the analysis of correct and
incorrect responses simultaneously creating Modified Module
Analysis - Partial (MMA-P). MMA-P applied the same methodology as
MMA, but to the partial correlation matrix rather than the
correlation matrix. This change allowed MMA-P to identify
incorrect communities as did MMA, but also fully correct
communities and communities that mixed correct and incorrect
responses. The incorrect communities identified were generally
consistent with those identified by MMA. The correct communities
generally followed those identified by Stewart {\it et al.}
\cite{stewart2018}; therefore, MMA-P productively identified both
correct solution structure and incorrect structure at the same
time. MMA-P also identified communities which mixed correct and
incorrect answers. These communities were productive in
understanding answering patterns where the correct answer was
possibly selected for the incorrect reason and in understanding
some of the gender unfairness identified by Traxler {\it et al.}.
The post-test reasoning of men and women was generally consistent.

\begin{acknowledgments}
This work was supported in part by the National Science Foundation
as part of the evaluation of improved learning for the Physics
Teacher Education Coalition, PHY-0108787.
\end{acknowledgments}

%

\end{document}